\documentclass[a4paper,fleqn]{cas-sc}

\usepackage[authoryear,longnamesfirst]{natbib}

\usepackage{bm}
\usepackage{booktabs}
\usepackage{multirow}
\usepackage{makecell}
\usepackage{graphicx}
\usepackage{subfigure}
\usepackage{tabularx}
\usepackage{times}
\usepackage{amsfonts}
 \usepackage{color}
 \usepackage{subfigure}
\usepackage{float}
\usepackage{graphicx}
\usepackage{url}
\usepackage{enumitem}
\usepackage{times}
\usepackage{url}
\usepackage{graphicx}
\usepackage{multirow}
\usepackage{booktabs}
\usepackage{color}
\usepackage{bm}
\setenumerate[1]{itemsep=0pt,partopsep=0pt,parsep=\parskip,topsep=5pt}
\setitemize[1]{itemsep=0pt,partopsep=0pt,parsep=\parskip,topsep=5pt}
\setdescription{itemsep=0pt,partopsep=0pt,parsep=\parskip,topsep=5pt}

\usepackage{setspace}
\usepackage{algorithm}  
\usepackage{algpseudocode}  
\usepackage{amsmath}
\usepackage{tabularray}
\usepackage{pifont}
\usepackage{tcolorbox}
\usepackage{booktabs} 
\usepackage{amssymb}
\usepackage{bbding}
\usepackage{svg}
\usepackage{bbding}
\usepackage{xcolor}
\usepackage{placeins}

\def\tsc#1{\csdef{#1}{\textsc{\lowercase{#1}}\xspace}}
\tsc{WGM}
\tsc{QE}

\begin{document}
\let\WriteBookmarks\relax
\def\floatpagepagefraction{1}
\def\textpagefraction{.001}

\shorttitle{}    

\shortauthors{}  

\title{Evidence, Logic, and Compliance: Multi-Agent Structured Graph Reasoning with Expert Arbitration for Medical Referral}

\author[scut,pu]{Qi Peng}
\author[scut]{Yi Cai}
\author[scut]{Jialin Cui}
\author[gd]{Tong Zhu}
\author[pu]{Yujuan Ding}
\author[gx]{Qingbao Huang}
\author[kcl]{Tao Wang}
\author[pu]{Jiayuan Xie}\cormark[1] 
\cortext[corresponding]{Corresponding author}
\ead{jiayuan.xie@polyu.edu.hk}
\author[pu]{Changmeng Zheng}\cormark[1] 
\ead{changmeng.zheng@polyu.edu.hk}  
\author[pu]{Qing Li}

\address[scut]{School of Software Engineering, South China University of Technology, Guangzhou, China}
\address[pu]{Department of Computing, Hong Kong Polytechnic University, Hong Kong, China}
\address[gd]{School of Computer Science and Network Engineering, Guangzhou University, Guangzhou China}
\address[gx]{School of Electrical Engineering, Guangxi University, Guangxi China}
\address[kcl]{Department of Biostatistics \& Health Informatics (BHI), King's College London, London}

\begin{abstract}
Medical referral (directing patients to the appropriate hospital department) is a complex decision-making process requiring the synthesis of multimodal data, including patient narratives, laboratory indicators, and radiology imaging. While Large Language Models (LLMs) have advanced medical dialogue systems, they struggle with real-world referral tasks due to two primary limitations: (1) Information Overload, where models fixate on high-frequency disease terms while overlooking subtle but critical urgency indicators; and (2) Unstructured Collaboration, where existing multi-agent frameworks rely on loose dialogue that leads to semantic drift and confirmation bias. To address these challenges, we introduce MASGR (Multi-Agent Structured Graph Reasoning), a framework that treats referral not as a classification task but as a structured graph construction problem. MASGR deploys specialized agents to extract evidence from distinct modalities and coordinates them through a clinical reasoning graph. This graph forces agents to establish explicit logical connections between conflicting evidence. Furthermore, we integrate a knowledge-guided arbitration mechanism that prioritizes patient safety rules over standard diagnostic classification. 
Extensive experiments on real-world medical records demonstrate that MASGR significantly outperforms state-of-the-art LLMs and existing multi-agent systems, particularly in complex cases requiring the balancing of chronic disease management and emergency intervention.
The AI contribution lies in the Multi-Agent Structured Graph Reasoning framework that transforms unstructured multi-agent dialogue into a verifiable logical graph construction. The engineering application is demonstrated through its deployment in a complex healthcare decision-making system to optimize the precision of complex medical referrals.
\end{abstract}





\maketitle

\section{Introduction}

\textbf{Medical referral}, the process of directing patients to the most appropriate hospital department, is a critical juncture in healthcare delivery \citep{anyanwu2015practice}. An accurate referral not only optimizes hospital resource allocation but, more importantly, ensures timely and life-saving treatment for patients with complex conditions. In real-world hospitals, this decision-making process is inherently multimodal and cognitively demanding \citep{kline2022multimodal,shaik2024survey}. It requires physicians to synthesize fragmented information from heterogeneous sources, ranging from subjective outpatient narratives to quantitative laboratory indicators, spatial radiology imaging, and microscopic pathology findings.

\begin{figure}[]
  \centering
  \includegraphics[width=0.5\linewidth]{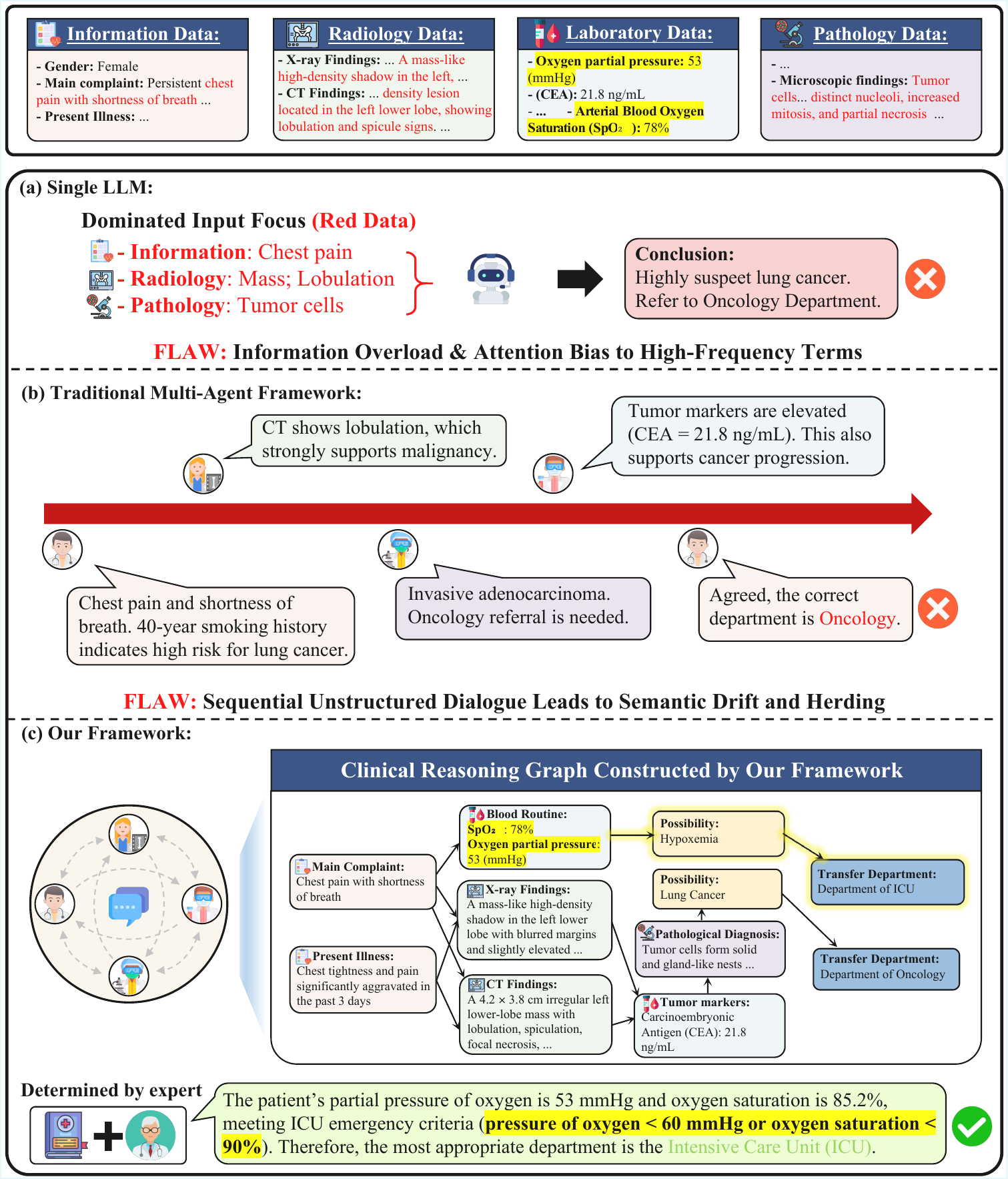}
  \caption{A comparison of our framework with existing decision-making paradigms for a patient with lung cancer complicated by severe hypoxemia.}
  \label{fig:intro}
\end{figure}

Recent advances in Large Language Models (LLMs) have shown great impressive capabilities across a wide range of medical applications \citep{liu2025improving,toma2023clinical,yang2025medaide,kumari2025novel,xu2026knowledge}. However, applying general-purpose LLMs to complex medical referral tasks presents significant challenges. \textit{First}, the \textbf{information overload} problem: when creating a prompt with all available patient data, single-agent LLMs often suffer from attention dispersion \citep{vishwanath2025medical,fan2025medodyssey}. As illustrated in Figure 1(a), a model may fixate on salient, high-frequency terms such as ``mass-like high-density shadow'' in radiology or ``tumor cells'' in pathology reports. Consequently, it may overlook subtle but critical urgency indicators (\textit{e.g.,} a drop in blood oxygen in lab results), which lead to technically correct disease identification but clinically disastrous referral decisions (\textit{e.g.,} sending a patient in respiratory failure to Oncology instead of the ICU). \textit{Second}, the lack of \textbf{structured reasoning}: existing solutions, including recent medical multi-agent frameworks \citep{tang2024medagents,kim2024mdagents,peng2024integration,yan2024clinicallab}, typically rely on natural language dialogue for collaboration. This unstructured interaction resembles a disorganized discussion; information is lost in translation, and there is no explicit verification of logical consistency \citep{ma2025medla}. As shown in Figure 1(b), agents may engage in ``herding'' behavior: seeing elevated tumor markers and CT lobulation, they mutually reinforce a diagnosis of invasive adenocarcinoma. This semantic drift allows the conversation to bypass the patient's critical hypoxemia entirely. Without a structured ``cognitive bridge'' connecting evidence to conclusions, the reasoning process remains opaque and prone to hallucination.

To address these limitations, we introduce the \textbf{M}ulti-\textbf{A}gent \textbf{S}tructured \textbf{G}raph \textbf{R}easoning (\textbf{MASGR}) framework. Inspiring from the workflow of a multidisciplinary team (MDT) of specialists \citep{taberna2020multidisciplinary}, our framework models the referral process not as a classification task, but as a structured graph construction and reasoning problem. The MASGR framework operates on three pillars:
\begin{itemize}
[itemsep=2pt,topsep=0pt,parsep=0pt,leftmargin=0.4cm]
    \item  \textbf{Domain-Specific Local Graph Construction:} Instead of a single overwhelmed model, we deploy specialized agents to process distinct data modalities (\textit{i.e.,} Outpatient, Lab, Radiology, and Pathology). Each agent acts as a domain expert, extracting evidence to construct a structured local reasoning graph. This ensures that subtle cues, such as the specific texture of a lung nodule in radiology or a critical drop in blood gas values, are captured with high fidelity before global integration.
    
    \item \textbf{Global Graph Co-Building via Structured Relations:} We replace unstructured dialogue with a collaborative cross-domain linking mechanism. Agents communicate by establishing explicit logical relationships between clinical evidence nodes: cascade (causal chains), syndrome (mutual reinforcement), exclusion (conflict identification), and context (conditional validity). This structured approach forces the system to align conflicting data, such as identifying a ``diagnostic discordance'' between a chronic tumor diagnosis and an acute respiratory crisis, into a coherent global reasoning graph, serving as a transparent investigation board.
    
    \item \textbf{Expert Knowledge Arbitration:} A correct diagnosis does not always equal a correct referral. To bridge the gap between medical facts and hospital operations, we introduce an expert knowledge arbitration mechanism. By referencing a curated knowledge base, an expert agent evaluates candidate reasoning graphs against three dimensions: evidence completeness, reasoning coherence, and knowledge compliance. For instance, as shown in Figure~\ref{fig:intro}(c), the system applies the strict compliance rule that ``pressure of oxygen $<$ 60 mmHg'' constitutes a medical emergency, correctly overriding the cancer diagnosis to prioritize the Intensive Care Unit (ICU).
\end{itemize}

We constructed a dataset based on real-world inpatient records from a tertiary hospital to evaluate our approach. Extensive experiments demonstrate that MASGR significantly outperforms state-of-the-art LLMs and existing medical multi-agent systems. Furthermore, ablation studies confirm that explicitly modeling logical relationships and integrating expert arbitration are crucial for handling complex cases.

\section{Related Work}

\subsection{Medical Large Language Models}
Large Language Models (LLMs) have been widely adopted in domains such as software development \citep{hou2024large,fan2023large}, fundamental sciences \citep{zhang2024geogpt}, and social simulation \citep{park2023generative}. More recently, they have shown great promise in the medical domains, with applications such as medical question answering \citep{singhal2025toward,lucas2024reasoning} and report generation \citep{liu2024bootstrapping}.
Medical LLMs typically follow two main development strategies: 
(1) training specialized models on domain-specific medical data \citep{wang2023chatcad}, and 
(2) adapting general-purpose LLMs through prompt engineering and retrieval-augmented generation (RAG) techniques \citep{wen2024mindmap}.
Initial efforts in medical LLMs centered on pre-training and fine-tuning models using domain-specific medical data. With the advent of large-scale general-purpose LLMs, however, training-free methods have gained attraction, which leverage both their inherent capabilities and external medical knowledge. For example, GPT-5 \citep{achiam2023gpt}, when guided by well-designed prompts, has demonstrated superior performance compared to specialized fine-tuned models such as Med-PaLM \citep{nori2023capabilities}.
Though LLMs have shown strong performance in the medical domain, they face limitations in complex scenarios involving multi-source medical data \citep{peng2024integration}. 
To address this, we propose a multi-agent framework where multiple LLMs collaboratively process diverse medical data, extract key information, and explicitly construct inter-data associations, improving the accuracy in complex medical scenarios.

\subsection{Multi-Agent Frameworks}
Previous studies have shown that multi-agent collaboration can significantly improve performance in complex tasks, such as large-scale code development \citep{hong2023metagpt}, strategic game playing \citep{zheng2024picture,liang2026multi}, and multi-robot coordination \citep{heuer2024benchmarking}. In recent years, this approach has been extended to the medical domain. For example, Tang et al. \citep{tang2024medagents} propose a Multidisciplinary Collaboration (MC) framework, where multiple agents engage in collaborative debate to reach medical diagnoses. Kim et al. \citep{kim2024mdagents} introduce MDAgents, a system that decomposes medical tasks into different complexity levels and assigns agent teams accordingly. These studies demonstrate the effectiveness of designing role-specific doctor agents to collaboratively perform medical decision-making.
However, existing multi-agent approaches often employ unstructured natural language interactions, causing topic drift or herd effects, which may lead to suboptimal referral decisions.
To address this issue, we propose a multi-agent reasoning framework. In this framework, multiple agents collaboratively extract key information from different types of medical data and finally integrate them into a reasoning graph dynamically. This reasoning graph models the causal relationships among symptoms, examinations, and diagnoses in clinical settings, improving inference accuracy and enhancing the interpretability of the final decision. 

\section{Data Construction}
\label{sec:data_construction}

\subsection{Dataset}
\subsubsection{Data Collection}
Since no publicly available dataset exists for evaluating large models in medical referral scenario, we collected a real-world dataset from a major general hospital.
We focused on inpatients during a selected time window and accessed the hospital's electronic medical record (EMR) system to retrieve all relevant clinical information from admission to department transfer.
The collected data include patient information, laboratory test results, imaging reports, pathology reports, and official referral records.
All files are indexed and organized using each patient’s unique hospitalization ID to ensure consistency and facilitate downstream modeling and analysis.

\subsubsection{Data processing}
After obtaining the raw data, we first structured the information of all patients. Specifically, we saved each patient's medical data type as the key and the corresponding result as the value in a JSON format file.
We then formulated the medical referral task as a multi-class classification problem.
To enable objective evaluation, we generated four candidate department options for each case. These included the ground truth departments from the transfer record and other randomly selected distractors from the hospital’s remaining departments.
The final structured JSON file for each patient contains demographic information, categorized medical test results, candidate department options, and the corresponding labels.

\begin{table}[]
\caption{Comparison with existing datasets. Our dataset covers all types of medical data.}
\label{tab:comparison}
\centering
\resizebox{0.75\textwidth}{!}{%
\begin{tabular}{l|cccc|cc}
\toprule
\textbf{Dataset} & 
\textbf{Patient Info} & 
\textbf{Radiology} & 
\textbf{Laboratory} & 
\textbf{Pathology} & 
\textbf{Avg. Words} & 
\textbf{Departments} \\
\midrule

PMC-VQA (2023)    & \XSolidBrush & \Checkmark & \XSolidBrush & \Checkmark & 9.96 &  2 \\
Slake (2021)       & \XSolidBrush & \Checkmark & \XSolidBrush & \XSolidBrush &  4.47  & 1\\
Path-VQA (2021)  & \XSolidBrush & \XSolidBrush & \XSolidBrush & \Checkmark & 6.23 & 1 \\
VQA-RAD (2018) & \XSolidBrush & \Checkmark & \XSolidBrush & \XSolidBrush & 6.80 & 1 \\
MedQA (2021)     & \Checkmark & \XSolidBrush & \Checkmark & \XSolidBrush &  116.18 & - \\
Open-XDDx (2025)     & \Checkmark & \XSolidBrush & \Checkmark & \XSolidBrush &  113.73 & 9 \\
MedThink-Bench (2025)     & \Checkmark & \XSolidBrush & \Checkmark & \Checkmark &  151.16 & 10 \\

\midrule
\textbf{Ours} & \Checkmark & \Checkmark & \Checkmark & \Checkmark & \textbf{1956.01} & \textbf{37} \\

\bottomrule
\end{tabular}
}
\end{table}

\begin{figure}[]
  \centering
    \includegraphics[width=0.7\textwidth]{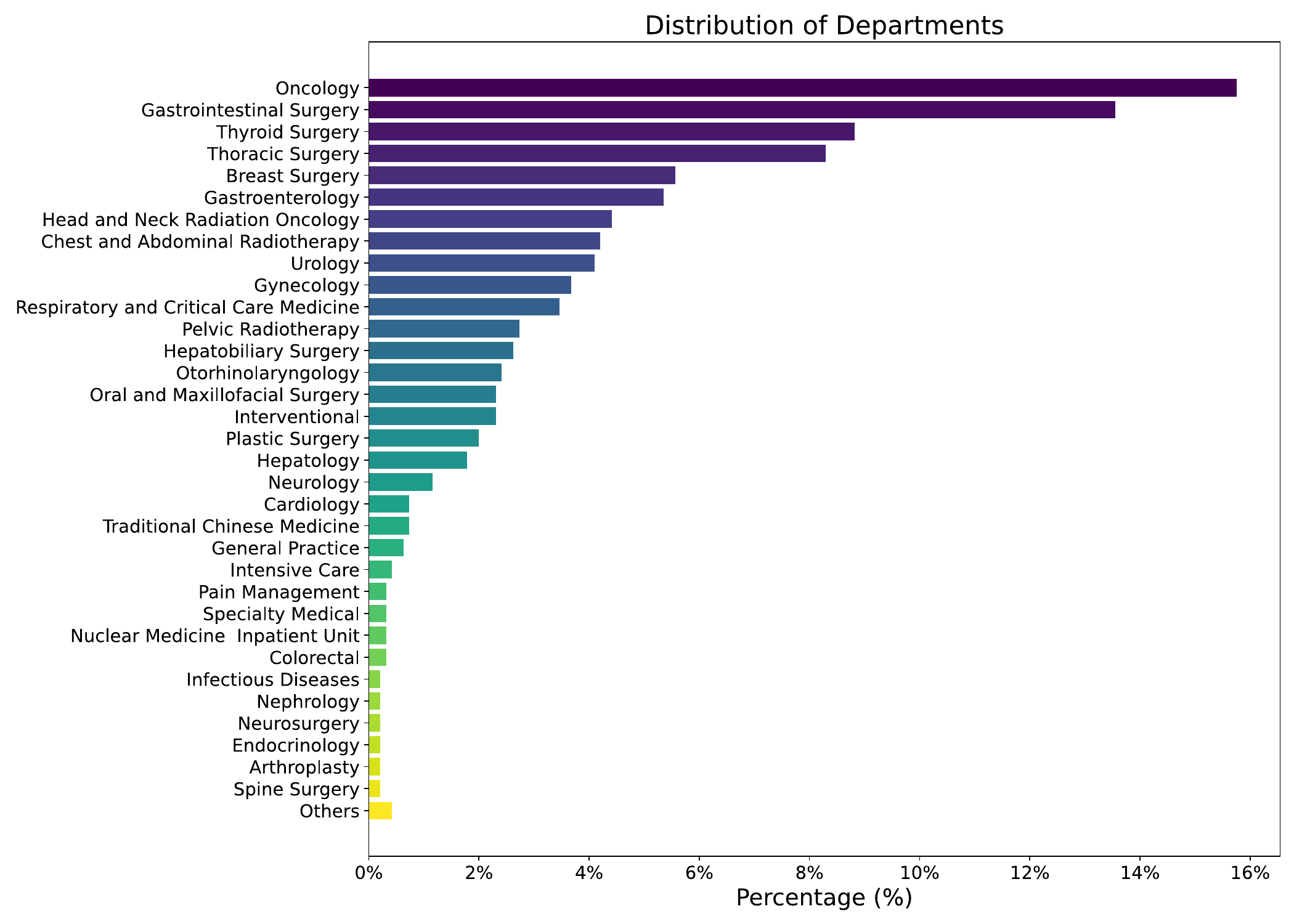}
  \caption{The distribution of departments in dataset.}
  \label{fig:department}
\end{figure}

\subsubsection{Data Statistic}
The final dataset consists of 952 patients, covering 37 clinical departments and 4,823 distinct disease types.
The top five most common diagnoses are hepatic cyst (2.8\%), hypertension (2.3\%), hypoproteinemia (1.7\%), type 2 diabetes mellitus (1.7\%), and chronic gastritis (1.6\%).
In terms of gender distribution, female patients account for 46.1\%, while male patients account for 53.9\%.
The three most frequently visited departments are oncology (15.76\%), gastrointestinal surgery (13.55\%), and thyroid surgery (8.82\%). The overall distribution of departments is shown in Figure~\ref{fig:department}.
Table~\ref{tab:comparison} compares our proposed dataset with existing benchmarks. We categorize medical data into four distinct types: Patient Info (e.g., chief complaint, present illness), Radiology (e.g., X-ray, CT reports), Laboratory (e.g., blood tests, urinalysis), and Pathology (e.g., biopsy findings). Additionally, the column ``Avg. Words'' indicates the average length of the input text for each dataset, and ``Departments'' means the number of departments.
As observed in the table, most existing datasets focus on limited data modalities. 
In contrast, our dataset achieves comprehensive coverage of all four data types. Moreover, each case in our dataset consistently contains all four modalities, whereas cases in existing datasets typically include only one or two types of examinations. This design leads to substantially more complex clinical scenarios, which require more advanced reasoning and decision-making capabilities to integrate heterogeneous medical evidence.


\subsection{Referral Knowledge Base}

To equip the expert agent with reliable domain knowledge and to ensure clinically grounded arbitration, we construct a referral knowledge base (RKB) that provides a minimally complete and interpretable framework for department-level referral reasoning. For each department, the RKB incorporates eight categories of structured knowledge: \textit{Department Overview} and \textit{Major Conditions and Diagnoses} delineate the clinical scope and diagnostic boundaries, enabling precise semantic alignment between patient-specific evidence and departmental expertise. \textit{Recommended Patient Transfer Criteria} and \textit{Cases Not Suitable for This Department} offer positive and negative referral constraints, ensuring logical consistency and preventing misclassification. \textit{Urgency Guidance} introduces emergency transfer indicators essential for identifying time-critical referral scenarios, while \textit{Interdepartmental Boundaries and Collaborations} captures cross-specialty interactions essential for handling ambiguous or overlapping presentations. Finally, \textit{Key Diagnostic Clues or Tests} and \textit{Typical Case Examples} provide evidence-based anchors that align model-generated reasoning chains with established clinical patterns.

To construct this knowledge base, we employ GPT-5 with carefully designed prompt templates tailored to each knowledge field. To minimize hallucinations and enforce factual grounding, we integrate MedRAG \citep{xiong2024benchmarking} as an external reference source. MedRAG retrieves authoritative clinical materials from resources such as PubMed\footnote{https://pubmed.ncbi.nlm.nih.gov/}, StatPearls\footnote{https://www.statpearls.com/} and Wikipedia\footnote{https://www.wikipedia.org/}, ensuring that the generated content remains objective, verifiable, and aligned with medical guidelines. 
For example, a partial prompt used for constructing \textit{Department Overview} field is:

\tcolorbox[colback=gray!20, colframe=gray!20, boxrule=0pt, sharp corners, left=2pt, right=2pt, top=2pt, bottom=2pt]
You are a medical knowledge extraction agent. Your task is to generate an accurate, concise, and guideline-aligned 
Department Overview for a given medical department.

......

You will be given:
(1) The target department name.
(2) A set of reference passages retrieved from MedRAG ......

Return the Department Overview in the following format: ......
\endtcolorbox

To rigorously validate the quality of the constructed RKB, we conducted a human evaluation involving three board-certified clinicians. We randomly sampled 10\% of the department entries and established a standardized evaluation protocol based on a 5-point Likert scale ($1=$ poor, $5=$ excellent). The experts assessed each entry across three dimensions: 
(1) \textbf{Clinical Accuracy}, measuring the factual correctness and absence of hallucinations in the generated medical content; 
(2) \textbf{Information Completeness}, evaluating whether the entry comprehensively covers all eight structured categories defined in our schema without omitting critical diagnostic criteria; and 
(3) \textbf{Referral Utility}, assessing the practical value and distinctiveness of the guidance for distinguishing inter-departmental boundaries. 
The evaluation yielded average scores of 4.8 for accuracy, 4.7 for completeness, and 4.7 for utility, demonstrating the high reliability of our construction pipeline. Furthermore, the Fleiss' Kappa score \citep{falotico2015fleiss} was calculated to be 0.76, indicating substantial inter-annotator agreement and confirming the robustness of the knowledge base as a trustworthy clinical reference.

\section{Problem Formulation}

\subsection{Task Input and Objective}

\textbf{Task Input:} 
Each case involves a \textit{multi-source patient record} denoted as $P = \{P_1, P_2, \ldots, P_k\}$, where each $P_i$ represents a specific information source, such as laboratory test results, imaging reports or prior medical history. These heterogeneous data sources collectively describe the patient's current clinical status.
The system operates over a predefined set of candidate medical departments 
$\mathcal{D} = \{Dept_1, Dept_2, \ldots, Dept_N\}$, encompassing common hospital divisions such as \textit{Cardiology}, \textit{Pulmonology}, and \textit{Emergency Medicine}.

\textbf{Task Objective:}  
Given the patient record $P$, the goal is to infer the most appropriate target department
\[
R_{\text{dept}} \in \mathcal{D},
\]
that can provide optimal diagnosis or treatment for the patient's current condition.

\subsection{Task Formulation: Reasoning Graph Construction}

To address this task, we re-formulate the medical referral task as a multi-stage, structured graph generation problem, defining the task as the following pipeline:

\textbf{Stage 1: Local Reasoning Graph Generation.}
This stage is executed by each domain agent $A_i \in \mathcal{A}$. The task for each agent $A_i$ is to independently analyze its designated data $P_i$ and construct a preliminary reasoning graph $\mathcal{G}_{local, i}$.
$$ \mathcal{G}_{local, i} = A_i(P_i) $$
Here, $\mathcal{G}_{local, i} = (\mathcal{V}_i, \mathcal{E}_i)$ comprises the evidence nodes $\mathcal{V}_i$ and their internal logical relationships $\mathcal{E}_i$. The collective output of this stage is a set of preliminary reasoning graphs $\{\mathcal{G}_{local, 1}, \mathcal{G}_{local, 2}, \ldots, \mathcal{G}_{local, N}\}$.

\textbf{Stage 2: Global Reasoning Graph Co-Building.}
This stage is executed by the multi-agent system $\mathcal{M}$ through collaboration among all agents $\mathcal{A}$. The task is to define a collaborative generation function $\Phi_{gen}$ that takes all preliminary graphs $\{\mathcal{G}_{local, i}\}$ as input. Through agent collaboration, this function constructs a set of candidate global reasoning graphs, $\mathbf{G}_{cand}$.
$$ \mathbf{G}_{cand} = \Phi_{gen}(\{\mathcal{G}_{local, 1}, \ldots, \mathcal{G}_{local, N}\}, \mathcal{M}) $$
In this formulation, $\mathbf{G}_{cand}$ represents the set of all plausible global graphs $\{\mathcal{G}_{global, j}\}$. Each candidate $\mathcal{G}_{global, j} = (\mathcal{V}_{global}, \mathcal{E}_{global})$ must be a logically coherent graph merging evidence from multiple sources. The key challenge is designing $\Phi_{gen}$ to effectively merge, link, and resolve the evidence and logic from multiple $\mathcal{G}_{local, i}$.

\begin{figure*}[]
  \centering
  \includegraphics[width=0.9\linewidth]{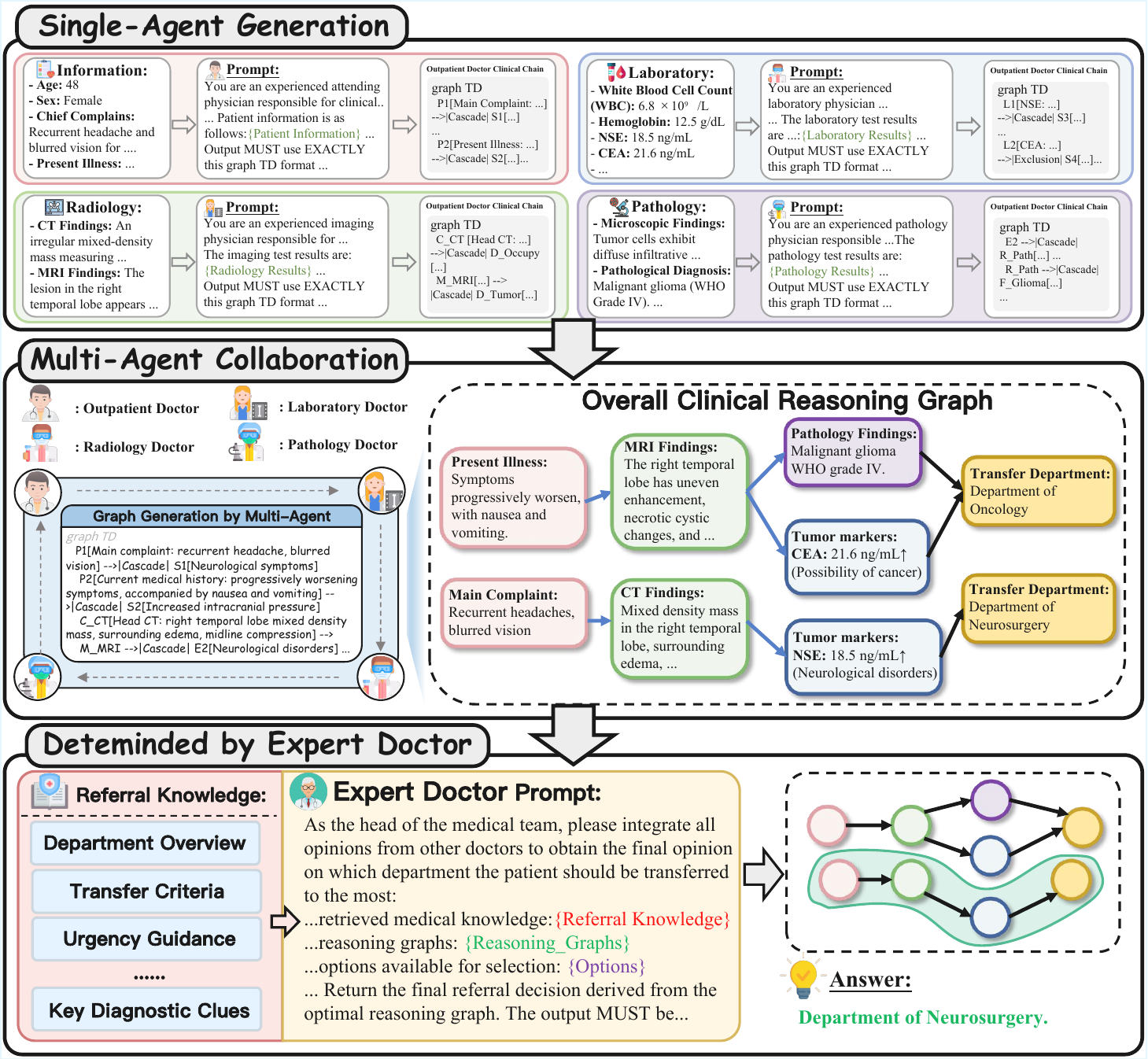}
      \caption{The overview of Multi-Agent Structured Graph Reasoning (MASGR) framework. The upper part is the process of single-agent reasoning graph generation. The middle part is multi-agent collaboration to correct and integrate the final reasoning graphs. The lower part is how our expert agent selects final department.}
  \label{fig:model}
\end{figure*}

\textbf{Stage 3: Expert-Knowledge Arbitration.}
This final stage is executed by the Expert Agent $A_{expert}$ augmented with an Expert Knowledge Base $\mathcal{K}_{expert}$. The objective is to evaluate the logical validity of the candidate graphs in $\mathbf{G}_{cand}$ and identify the most reasonable reasoning chain, denoted as $\mathcal{G}_{optimal}$.
$$ \mathcal{G}_{optimal} = \Phi_{arb}(\mathbf{G}_{cand}, \mathcal{K}_{expert}) $$
Here, $\Phi_{arb}$ represents the arbitration process where the expert agent filters candidates based on domain knowledge. Finally, the specific department recommendation $R_{dept}$ is directly derived from the diagnostic path within this optimal reasoning chain.
$$ R_{dept} = \Psi_{decide}(\mathcal{G}_{optimal}) $$
where $\Psi_{decide}$ maps the confirmed diagnosis path to the target department $R_{dept} \in \mathcal{D}$.

\section{Methodology}
To address the medical referral task, we propose a Multi-Agent Clinical Reasoning Graph Construction Framework, which models clinical decision-making as a multi-stage graph construction process. As illustrated in Figure~\ref{fig:model}, the framework dynamically constructs and aggregates reasoning structures across heterogeneous medical domains. It operates in three sequential stages: (1) Local Reasoning Graph Construction by single agent, (2) Global Reasoning Graph Co-Building through multi-agent collaboration, and (3) Expert Knowledge Arbitration for final decision.




\subsection{Stage 1: Local Reasoning Graph Construction}

Considering real-world clinical workflows, medical data can be categorized into four main types: \textit{outpatient records}, \textit{laboratory tests}, \textit{radiology reports}, and \textit{pathology reports}. Accordingly, we define four domain agents:
\[
A = \{A^{out}, A^{lab}, A^{rad}, A^{pat}\}
\]
Each agent $A^i$ is responsible for interpreting data from its respective domain $P_i$ and constructing a structured local reasoning graph.
Formally, for each agent $A^i$, we define:
\[
\mathcal{G}_{local, i} = A^i(P_i) = (\mathcal{V}_i, \mathcal{E}_i)
\]
where $\mathcal{G}_{local, i}$ is the local reasoning graph, $\mathcal{V}_i$ denotes the set of extracted clinical evidence nodes, and $\mathcal{E}_i$ represents the logical relationships among them.
To ensure consistent and controllable reasoning behavior across heterogeneous domains, each agent $A^i$ is guided by a structured prompt that specifies how raw clinical inputs are abstracted into evidence nodes, how intermediate clinical inferences are formed, and how referral-oriented conclusions are represented in a graph form.  
The prompt template shared by all agents is:

\tcolorbox[colback=gray!20, colframe=gray!20, boxrule=0pt, sharp corners, left=2pt, right=2pt, top=2pt, bottom=2pt]
{
You are \{domain\_specific\} physician responsible for ... and generate a reasoning graph to recommend the most appropriate specialty department. \\
... \\
The domain-specific data are as follows:\\
\{domain\_specific\_data\}\\
Generate a department referral reasoning graph by following these principles:\\
1. Start with abnormal test results and their deviations from reference ranges. \\
\dots \\
Output MUST use EXACTLY this \textbf{graph TD} format (with line breaks between nodes) and the content is for
reference only:\\
\dots
}
\endtcolorbox

\begin{figure}[]
  \centering
  \includegraphics[width=0.5\linewidth]{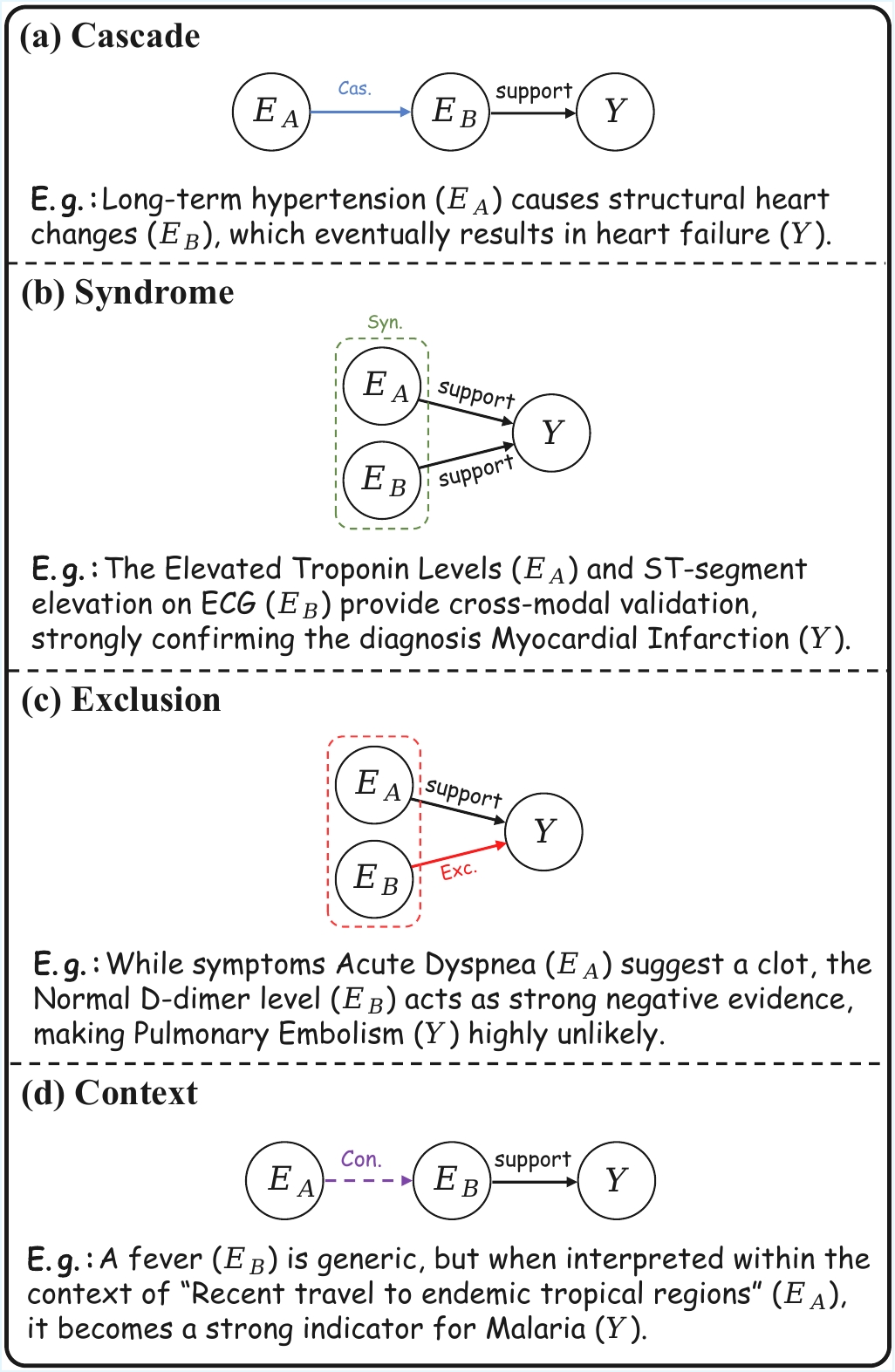}
      \caption{The four logical relationships among clinical evidence. (a) Cascade: $E_A$ leads to $E_B$, which ultimately supports outcome $Y$. (b) Syndrome: $E_A$ and $E_B$ jointly point to outcome $Y$. (c) Exclusion: $E_B$ conflicts with outcome $Y$, which was derived from $E_A$. (d) Context: $E_B$ can only lead to outcome $Y$ when $E_A$ serves as a precondition.}
  \label{fig:relationship}
\end{figure}

\subsection{Stage 2: Global Reasoning Graph Building}

This stage aims to integrate the local graphs into the global reasoning graphs through multi-agent collaboration. The process consists of two key mechanisms: 1) Cross-Domain Linking, where agents exchange and reconcile their reasoning structures through relationship-aware updates, and 2) Graph Aggregation, where the converged local graphs are merged into logically coherent global graph.

\subsubsection{Cross-Domain Linking}

At iteration $t$, each agent $A^i$ maintains its current reasoning graph $\mathcal{G}_{local, i}^{(t)} = (\mathcal{V}_i^{(t)}, \mathcal{E}_i^{(t)})$. During each round of collaboration, every agent interacts with all others, updating its graph:
\[
\mathcal{G}_{local, i}^{(t+1)} = \Phi_i\Big(\mathcal{G}_{local, i}^{(t)}, \{\mathcal{G}_{local, i}^{(t)} \mid j \neq i\}\Big),
\]
where $\Phi_i(\cdot)$ denotes the graph update operator of agent $A^i$.

To ensure that cross-agent interactions lead to consistent and interpretable structural updates, the operator $\Phi_i(\cdot)$ is implemented via a structured interaction prompt, which instructs each agent to revise its local reasoning graph by incorporating new evidence and feedback from other agents while preserving explicit reasoning relations. 
A generic interaction prompt shared by all agents is summarized below:

\tcolorbox[colback=gray!20, colframe=gray!20, boxrule=0pt, sharp corners, left=2pt, right=2pt, top=2pt, bottom=2pt]
{
As the \{doctor\_type\} doctor, please generate an UPDATED reasoning graph based on the original assessment, new medical data, and feedback from other doctors. \\
Output MUST use EXACTLY this \textbf{graph TD} format (with line breaks between nodes) and the content is for reference only:\\
...\\
Use specific relationship types between nodes: \\
...\\
Please ensure that the output strictly follows the above format. Avoid any additional text or explanations outside the graph structure. 
}
\endtcolorbox

Let $\sigma: \bigcup_i \mathcal{V}_i^{(t)} \rightarrow \mathcal{C}$ denote the mapping from agent-specific evidence to a shared clinical concept space $\mathcal{C}$.
For each iteration $t$, agent $A_i$ updates its reasoning structure based on the following relational mappings:

\textbf{(i) Cascade:}
This relation captures a causal trajectory where an antecedent condition ($E_A$) triggers a subsequent pathological state ($E_B$), which directly leads to the final diagnosis ($Y$).
If two agents discover sequential or causal chains across their domains:
\[
e_a^{(i)} \xrightarrow{\text{Cas.}} e_b^{(i)}, \quad e_c^{(j)} \xrightarrow{\text{Cas.}} e_d^{(j)}
\]
and their mapped concepts satisfy $\sigma(e_a^{(i)})=\sigma(e_c^{(j)}), \sigma(e_b^{(i)})=\sigma(e_d^{(j)})$, then a unified progressive edge is established:
\[
[e_a] \xrightarrow{\text{Cas.}} [e_b] \in \mathcal{E}_i^{(t+1)} \cup \mathcal{E}_j^{(t+1)}.
\]
Here, the notation $[e_a] \xrightarrow{\text{Cas.}} [e_b]$ denotes that a pathological evolution is formed between the semantic equivalence classes $[e_a]$ and $[e_b]$, which represent concept-level evidence clusters shared across agents.  
The edge set $\mathcal{E}_i^{(t+1)} \cup \mathcal{E}_j^{(t+1)}$ indicates that this newly unified edge is jointly integrated into the updated reasoning graphs of agents $A_i$ and $A_j$ at iteration $(t+1)$, thus synchronizing their causal reasoning structures within the multi-agent system.

\textbf{(ii) Syndrome:}
This relation describes a consensus mechanism where independent clinical findings ($E_A$ and $E_B$) mutually reinforce each other to heighten the confidence of the diagnosis ($Y$).
If multiple agents provide complementary evidence $\{e_1^{(i)}, e_2^{(j)}, \dots, e_m^{(k)}\}$ supporting the same conclusion $\sigma(v)$:
\[
\sigma(e_1^{(i)}), \sigma(e_2^{(j)}), \dots, \sigma(e_m^{(k)}) \to \sigma(v),
\]
then a conjunctive structure is formed in the next update:
\[
\{[e_1], [e_2], \dots, [e_m]\} \xrightarrow{\text{syn.}} [v] \in \mathcal{E}_i^{(t+1)}.
\]

\textbf{(iii) Exclusion:}
This relation represents a reasoning conflict where positive evidence ($E_A$) suggests the diagnosis ($Y$), while contradictory evidence ($E_B$) acts as a counter-indicator to rule it out.
If cross-agent evidence pairs are known to be mutually exclusive according to the medical ontology $\mathcal{X}$:
\[
\{\sigma(e_a^{(i)}), \sigma(e_b^{(j)})\} \in \mathcal{X},
\]
then a contradiction edge is added:
\[
[e_a] \perp [e_b] \in \mathcal{E}_i^{(t+1)} \cup \mathcal{E}_j^{(t+1)}.
\]

\textbf{(iv) Context:}
This relation indicates that the diagnostic validity of a specific symptom ($E_B$) towards the diagnosis ($Y$) is conditional upon or amplified by the patient's background context ($E_A$).
When the validity of an evidence $e_b^{(i)}$ depends on a prerequisite condition $e_a^{(j)}$:
\[
e_a^{(j)} \xrightarrow{\text{Con.}} e_b^{(i)},
\]
the conditional dependency is preserved:
\[
[e_a] \xrightarrow{\text{Con.}} [e_b] \in \mathcal{E}_i^{(t+1)}.
\]

After $T$ iterations, all agents reach consensus, yielding a converged set of reasoning graphs $\{\mathcal{G}_i^{*}\}$:
\[
\mathcal{G}_{local, i}^{(T)} = \mathcal{G}_{local, i}^{(T-1)} = \mathcal{G}_{local, i}^{*}.
\]

\subsubsection{Graph Aggregation}




The global reasoning graphs are obtained via an aggregation operator $\Gamma(\cdot)$ that merges the converged local graphs into a set of candidate global reasoning graphs:
\[
\mathbf{G}_{cand} = \Gamma\Big(\{\mathcal{G}_{local, i}^{(T)}\}_{i=1}^4\Big) = \{\mathcal{G}_{global, 1}, \dots, \mathcal{G}_{global, M}\}
\]
Each candidate $\mathcal{G}_{global, k} \in \mathbf{G}_{cand}$ represents a specific, logically connected diagnostic pathway derived from the multi-agent consensus. This set $\mathbf{G}_{cand}$ serves as the input for the subsequent Expert Knowledge Arbitration stage.

\subsection{Stage 3: Expert Knowledge Arbitration}

In this final stage, we employ an Expert Knowledge Arbitration mechanism to identify the most clinically valid reasoning path from the set of candidate global graphs $\mathbf{G}_{cand}$. 

The arbitration process, formalized as function $\Phi_{arb}$, evaluates each candidate graph along three critical dimensions to ensure high-quality clinical decision-making:

\begin{itemize}
    \item \textbf{Evidence Completeness:} Verifying whether the reasoning graph sufficiently incorporates all key clinical evidence from the patient case, including symptoms, medical history, diagnostic tests, and examination results.
    \item \textbf{Reasoning Coherence:} Assessing whether the causal and logical structure of the graph is medically reasonable, internally consistent, and free from logical conflicts.
    \item \textbf{Knowledge Compliance:} Ensuring that the diagnostic path and referral conclusions strictly conform to established clinical guidelines and expert rules defined in the Expert Knowledge Base ($\mathcal{K}_{expert}$).
\end{itemize}

To operationalize $\Phi_{arb}$, the expert agent is guided by a structured arbitration prompt, which instructs it to semantically validate candidate reasoning graphs against expert knowledge and select the optimal referral outcome. 
The generic prompt used by $A_{expert}$ is summarized below:

\tcolorbox[colback=gray!20, colframe=gray!20, boxrule=0pt, sharp corners, left=2pt, right=2pt, top=2pt, bottom=2pt]
{
As the head of the medical team, please integrate all opinions from other doctors to obtain the final opinion on which department the patient should be transferred to the most.\\
Inputs:\\
1. Patient case information: \{patient\_case\}\\
2. Retrieved expert medical knowledge: \{retrieved\_info\}\\
3. Reasoning graphs: ...
}
\endtcolorbox
\tcolorbox[colback=gray!20, colframe=gray!20, boxrule=0pt, sharp corners, left=2pt, right=2pt, top=2pt, bottom=2pt]
{
... \\
Arbitration principles:\\
-- Evaluate each reasoning graph for evidence completeness, reasoning coherence, and knowledge compliance.\\
-- ...\\
Output requirements:\\
Return the final referral decision derived from the optimal reasoning graph.\\
The output MUST be ...
}
\endtcolorbox

Through this arbitration process, the system filters out hallucinatory or incomplete reasoning paths and outputs the optimal reasoning graph:
\begin{equation}
\mathcal{G}_{optimal} = \Phi_{arb}(\mathbf{G}_{cand}, \mathcal{K}_{expert})
\end{equation}

Finally, the department recommendation is determined by traversing the diagnostic path of the verified optimal graph. 
We define a decision function $\Psi_{decide}$ that maps the conclusion of the reasoning chain to a specific medical department:
\begin{equation}
R_{dept} = \Psi_{decide}(\mathcal{G}_{optimal}), \quad R_{dept} \in \mathcal{D}.
\end{equation}

\section{Experiment}

\subsection{Experimental Settings}
\subsubsection{Dataset}
Considering there is no dataset for complex medical referral task, we first construct a corresponding dataset based on inpatient records from a real-world hospital, which is mentioned in Section~\ref{sec:data_construction}. This dataset includes all relevant information prior to patient department transfer, such as basic patient information (e.g., sex, age, present illness), laboratory indicators (e.g., blood tests), radiology findings (e.g., X-ray, MRI), and pathology findings (e.g., frozen pathology). It also contains records of patient department transfers.
The dataset covers 952 patients across 37 departments and includes data from various medical examination modalities with temporal information.

\subsubsection{Baseline Methods}

We compare our framework with the following methods:

\begin{itemize}

\item \textbf{Large Language Models:} Considering the advancements of general-purpose LLMs in medical domain, we select several models for comparison, including GPT-5 \citep{achiam2023gpt}, Qwen-Turbo \citep{team2024qwen2}, Glm-4-Flash \citep{glm2024chatglm},
DeepSeek-V3 \citep{liu2024deepseek}, and DeepSeek-R1 \citep{guo2025deepseek}.

\item \textbf{Prompt-Based Methods:} We compare against prompt-enhanced methods for LLMs, such as chain-of-thought (CoT) \citep{wei2022chain}, tree-of-thought (ToT) \citep{yao2023tree}, and self-consistency \citep{wang2022self}.

\item \textbf{Multi-Agent Systems:} We also compare with medical multi-agent frameworks, such as MedAgents \citep{tang2024medagents}, MDAgents \citep{kim2024mdagents}, MMA \citep{peng2024integration} and ToR \citep{peng2025tree}. These frameworks are based on multi-agent collaboration to complete medical tasks.

\end{itemize}

\subsubsection{Experimental Details}
For Large Language Models, we directly use the corresponding API interfaces from the respective platforms, including OpenAI\footnote{https://openai.com/}, 
Qwen\footnote{https://chat.qwen.ai/}, ChatGLM \footnote{https://chatglm.cn/} and DeepSeek\footnote{https://www.deepseek.com/}.
For prompt-based methods and multi-agent systems, we use DeepSeek-V3 as the base model to ensure a fair comparison. Similarly, we access the API interfaces provided by the DeepSeek platform to implement these methods.
All methods are executed in a zero-shot setting.

\subsubsection{Evaluation Metrics}
In this study, we select both objective and subjective metrics to evaluate proposed methods. 
For objective evaluation, we model the task as a multi-class classification problem, where the goal is to select appropriate departments for patient based on all available medical examination data. Therefore, we use precision, recall, and F1 score as objective evaluation metrics to assess the performance of different methods.
For subjective evaluation, we assess the clinical reasoning quality of outputs from different methods. Specifically, two evaluation dimensions are defined:

\begin{enumerate}[leftmargin=0.45cm]
    \item \textbf{Evidence Completeness (EC)}:
    \begin{itemize}
        \item \textbf{Definition:} This metric evaluates whether the reasoning graph sufficiently incorporates all key clinical evidence from the patient case.
        \item \textbf{Scoring Guidelines (0-5 points):}
        \begin{itemize}[leftmargin=0.35cm]
            \item \textbf{5.0 points:} The graph captures all essential clinical findings. Every key piece of evidence required for the diagnosis is explicitly represented and correctly categorized.
            \item \textbf{4.0 points:} Includes the vast majority of core evidence. Only negligible details are omitted, which does not affect the final clinical decision.
            \item \textbf{3.0 points:} Captures the primary complaint and major symptoms but misses supporting diagnostic details. The basis for diagnosis is present but not robust.
            \item \textbf{2.0 points:} Misses significant clinical evidence or introduces irrelevant ``noise'' nodes that distract from the main diagnosis. The graph fails to reflect the complexity of the patient's condition.
            \item \textbf{1.0 points:} Fails to incorporate essential clinical data or contains hallucinations (evidence not present in the source text). The graph is medically insufficient.
        \end{itemize}
    \end{itemize}

    \item \textbf{Reasoning Coherence (RC)}:
    \begin{itemize}
        \item \textbf{Definition:} This metric assesses whether the logical connections among different clinical evidence are medically reasonable and consistent with established diagnostic knowledge.
        \item \textbf{Scoring Guidelines (0-5 points):}
        \begin{itemize}[leftmargin=0.35cm]
            \item \textbf{5.0 points:} The reasoning chain is logically flawless and explicitly causal. All deductions align perfectly with standard clinical guidelines, offering a clear interpretability path.
            \item \textbf{4.0 points:} The reasoning is generally sound and medically plausible. There may be minor logical gaps or implicit steps, but the overall diagnostic direction is correct and justifiable.
            \item \textbf{3.0 points:} The logical path is traceable but lacks depth. The connection between evidence and diagnosis relies on weak justifications or generic associations rather than specific pathological reasoning.
            \item \textbf{2.0 points:} Contains noticeable logical contradictions or significant leaps (non sequiturs). The graph connects unrelated evidence to conclusions without valid medical basis.
            \item \textbf{1.0 points:} The reasoning is medically invalid, dangerous, or completely unstructured. The relationships between nodes violate basic medical principles.
        \end{itemize}
    \end{itemize}

\end{enumerate}

Specifically, three experienced clinicians independently assessed each generated reasoning graph based on the two dimensions. Scoring is precise to one decimal place, and the final score for each dimension is computed as the average rating across all evaluators.

\subsection{Main Results}

\begin{table*}[t]
\footnotesize
\centering
\caption{Overall performance of different methods, including precision, recall and F1. \textbf{EC} means Evidence Completeness and \textbf{RC} means Reasoning Coherence. The best results are marked in \textbf{bold}. Statistical significance is evaluated using a paired t-test, and improvements are considered significant at $p < 0.01$.
}
\label{tab:results}
\resizebox{\textwidth}{!}{\begin{tabular}{llccccccc}
\toprule
\textbf{Category} & \textbf{Methods} & \textbf{Precision(\%)} & \textbf{Recall(\%)} & \textbf{F1(\%)} & \textbf{EC} & \textbf{RC}   \\
\midrule
\multirow{4}{*}{LLMs} & GPT-5 (2025) \citep{achiam2023gpt} & 85.0 & 95.1 & 89.7 & 3.83 & 3.4  \\
 & Qwen-Turbo (2024) \citep{team2024qwen2} & 82.9 & \textbf{97.0} & 89.4 & 3.8 & 3.36   \\ 
 & Glm-4-Flash (2024) \citep{glm2024chatglm} & 65.1 & 95.0 &  77.2 & 3.43 & 3.1  \\ 
 & DeepSeek-V3 (Liu et al. 2024) \citep{liu2024deepseek} & 82.9 & 93.8 & 88.0 & 3.8 & 3.33  \\
 & DeepSeek-R1 (Guo et al. 2025) \citep{guo2025deepseek} & 74.6 & 91.0 & 81.9 & 4 & 3.5  \\
\midrule
\multirow{4}{*}{Prompt-Based}
 & Self-Consistency (Wang et al. 2022) \citep{wang2022self}& 34.2 & 89.3 & 49.4 & 3.5 & 3.1  \\
 & CoT (Wei et al. 2022) \citep{wei2022chain} & 87.3 & 92.9 & 90.0 & 3.87 &  3.5  \\ 
 & ToT (Yao et al. 2023) \citep{yao2023tree} & 52.9 & 81.0 & 64.0  & 3.47 &  3.17  \\
 
\midrule
\multirow{3}{*}{Multi-Agent} 
  & MedAgents (Tang et al. 2024) \citep{tang2024medagents} & 90.9 & 89.7 & 90.3 & 4.1 & 3.47 \\ 
  & MDAgents (Kim et al. 2024) \citep{kim2024mdagents}  & 88.2 & 89.2 & 88.7 & 4.1 & 3.5 \\ 
  & MMA (Peng et al. 2024) \citep{peng2024integration} & 90.4 & 92.5 & 91.4 & 4.47 & 3.8  \\ 
  & ToR (Peng et al. 2025) \citep{peng2025tree} & 91.8 & 93.3 & 92.6 & 4.3 & 3.83  \\
\midrule
Our & Multi-Agent Structured Graph Reasoning (MASGR) & \textbf{95.2} & 95.4  & \textbf{95.3} & \textbf{4.57} & \textbf{4.2}  \\
\bottomrule
\end{tabular}}
\end{table*}

\begin{table*}[t]
\centering
\caption{Ablation study on different Agents. ``$\checkmark$'' means presence of the doctor agent and ``$\times$'' means absence of the doctor agent.}
\label{tab:abla_role}
\begin{tabular}{cccccccc}
\toprule
\textbf{Outpatient} & \textbf{Laboratory} & \textbf{Radiology} & \textbf{Pathology} & \textbf{Expert} & \textbf{P.(\%)} & \textbf{R.(\%)} & \textbf{F1(\%)} \\
\midrule
$\checkmark$ & $\times$ & $\times$ & $\times$ & $\times$ & 82.9 & 93.8 & 88.0 \\
$\checkmark$ & $\checkmark$ & $\times$ & $\times$ & $\times$ & 83.1  & 95.0 & 88.7 \\ 
$\checkmark$ & $\checkmark$ & $\checkmark$ & $\times$  & $\times$ & 89.3 & \textbf{97.0} & 93.0 \\ 
$\checkmark$ & $\checkmark$ & $\checkmark$ & $\checkmark$ & $\times$ & 90.3 & 96.0  & 93.1   \\ 
$\checkmark$ & $\checkmark$ & $\checkmark$ & $\checkmark$ & $\checkmark$ & \textbf{95.2} & 95.4  & \textbf{95.3}   \\
\bottomrule
\end{tabular}
\end{table*}


As shown in Table~\ref{tab:results}, we compared the performance of LLMs, prompt-based methods, multi-agent methods, and our framework using the evaluation metrics of precision, recall, and F1. We present the following key findings:

(i) For complex medical tasks involving different types of medical data, multi-agent methods outperform most of LLMs and prompt-based methods. Specifically, the F1 score of MedAgents is 90.3\%, whereas DeepSeek-R1 and CoT have F1 scores of 81.9\% and 90.0\%, respectively. Unlike traditional medical tasks, this task involves multiple types of medical data, which need to be simultaneously analyzed and correlated to determine the most appropriate department for transfer. As for LLM methods, they face the challenge of processing multiple medical data types from different modalities simultaneously, which can cause the model's attention to be dispersed and reduce its ability to effectively focus on the most relevant information. This leads to a decrease in overall performance. In contrast, in the multi-agent framework, each agent focuses on its specific task without interference from other medical data. Through collaboration, agents provide consistent diagnostic results, allowing for more comprehensive analysis and a more complete clinical reasoning output.

(ii) Our framework outperforms existing multi-agent systems. The F1 score of MASGR is 95.3\%, compared to the F1 score of 91.4\% for MMA. Unlike other multi-agent frameworks, our method proposes a collaborative multi-agent approach that constructs a clinical reasoning graph, associating key clinical information from different types of medical data, which can better capture and analyze complex medical data, improving reasoning accuracy. Additionally, we introduce an expert agent with clinical referral knowledge, which provides essential information on departmental functions, transfer criteria, and urgency guidance, thereby facilitating accurate referral decisions.


(iii) For the subjective evaluation metrics, our framework achieved the highest scores in both Evidence Completeness (EC) and Reasoning Coherence (RC). Specifically, the proposed method obtained an EC score of 4.57 and an RC score of 4.20, outperforming GPT-5 by 0.74 and 0.80, respectively, and exceeding the best-performing multi-agent baseline (ToR) by 0.27 and 0.37. These results demonstrate that, compared with existing single-LLM and multi-agent approaches, our method captures more comprehensive clinical evidence and establishes more coherent relationships among the evidence to support accurate decision-making. This capability is particularly crucial in real-world medical scenarios, as it enables the generation of more trustworthy reasoning chains, thereby offering high value for clinical decision support systems.

\subsection{Ablation Study}

\begin{figure*}[]
  \centering
  \includegraphics[width=0.9\linewidth]{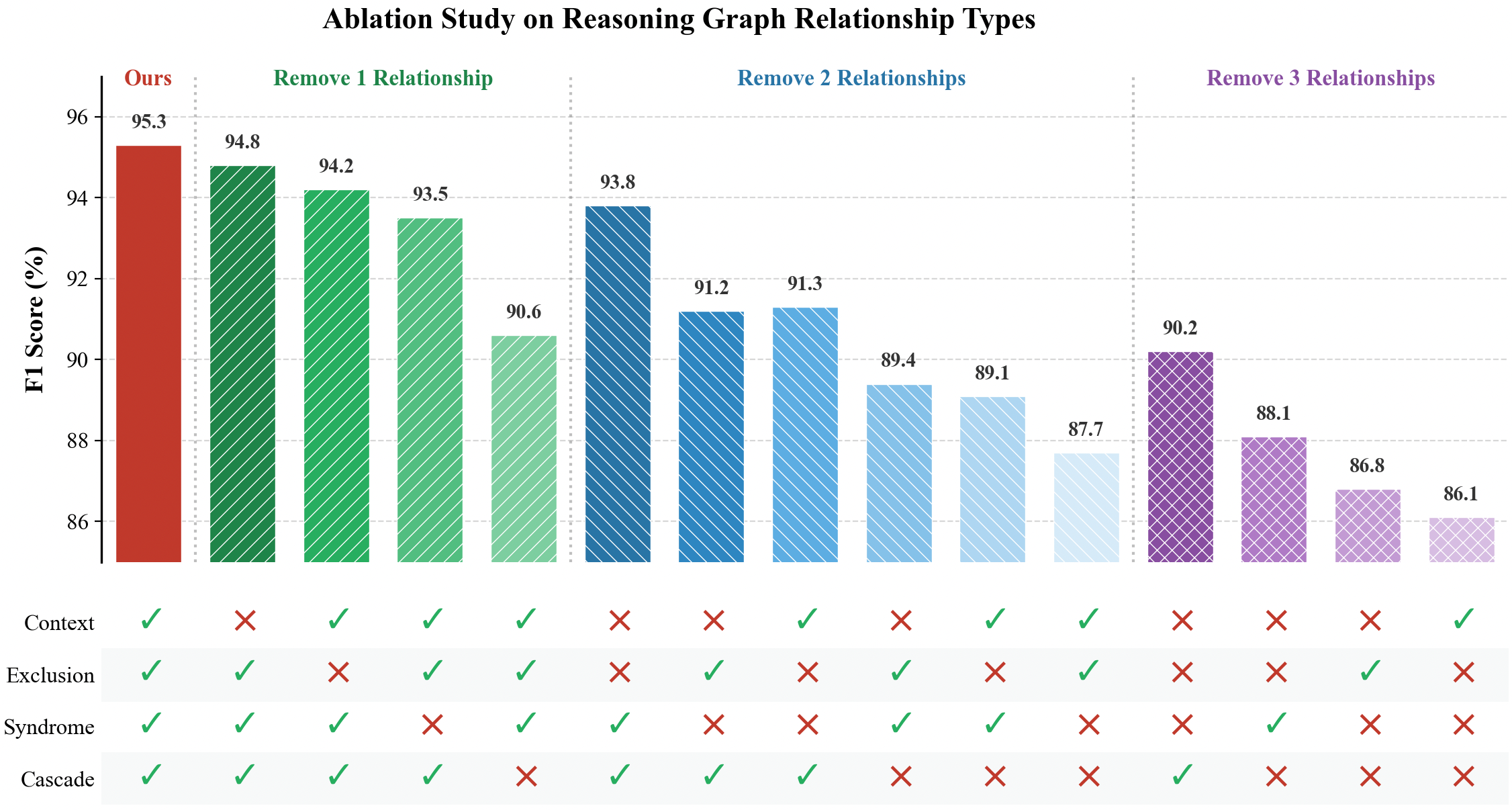}
  \caption{The ablation experimental result on four relationship types. The vertical axis represents the F1 score, while the horizontal axis illustrates the relationship types included in the reasoning graph. A relationship matrix is used for visualization, where green $\checkmark$ indicates that a specific relationship type is incorporated and red $\times$ denotes its exclusion.}
  \label{fig:four_type_ablation}
\end{figure*}

\begin{figure}[]
  \centering
  \includegraphics[width=0.5\linewidth]{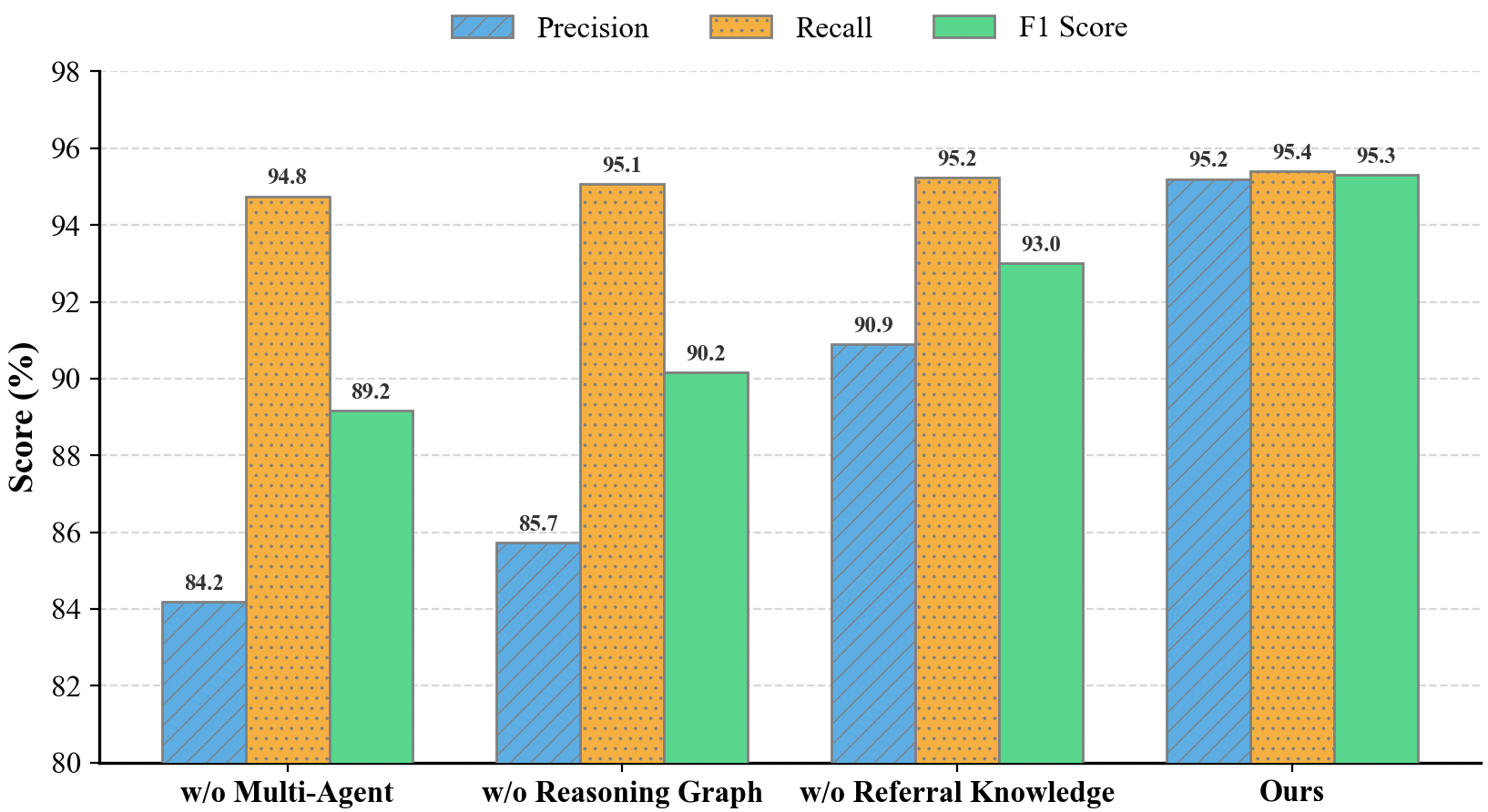}
  \caption{Ablation study on different mechanism or module. ``w/o'' means our framework without a certain part.}
  \label{fig:abla_module}
\end{figure}

\begin{figure}[]
  \centering
  \includegraphics[width=0.5\linewidth]{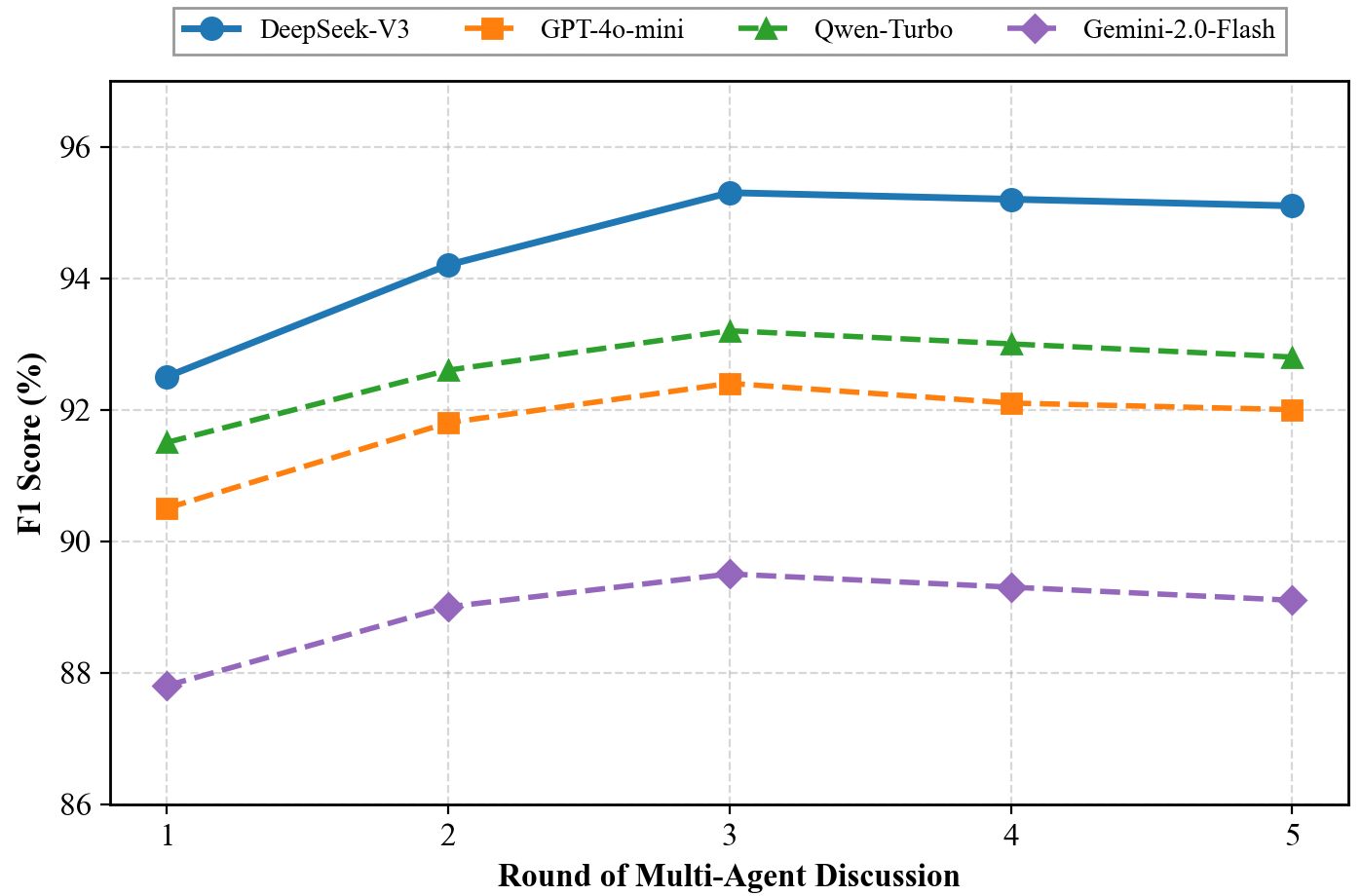}
  \caption{Performance comparison of different backbone models across varying discussion rounds. The plot illustrates the F1 score trends of DeepSeek-V3, Qwen-Turbo, GPT-4o-mini, and Gemini-2.0-Flash as the number of multi-agent discussion rounds increases from 1 to 5.}
  \label{fig:multi_round}
\end{figure}

To explore the impact of different agents on the final results, we conduct the ablation study shown in Table~\ref{tab:abla_role}. In this experiment, different agents were excluded from the system to observe the changes in precision, recall, and F1 score.
When a specific doctor agent is excluded, the corresponding medical data or responsibility are assigned to the outpatient doctor agent.
We made the following findings:

(i) \textbf{As the number of agents increases, the performance improves.} Specifically, when the laboratory doctor is introduced, the F1 score increases by 0.7\%. Further improvements are seen when the radiology and pathology agents are added, with F1 scores increasing by 4.3\% and 0.1\%, respectively. After involving the expert doctor agent, the F1 scores continuely increases by 2.2\%.
Further analysis reveals that when more agents are introduced, the data from different domains are handled by their respective agents, allowing for more precise extraction of key information from each data type, which improves the accuracy of the clinical reasoning graph.


(ii) \textbf{When the laboratory, radiology, pathology, and expert doctors are excluded, our framework reduces to a single LLM system}, resulting in an F1 score of 88.0\%. When faced with complex medical reasoning tasks, a single LLM must process multiple types of medical data simultaneously, which can cause the model's attention to shift and lead to the omission of key information from multi-source medical data, thus decreasing model performance.

To explore the impact of different modules on our framework, we conduct the corresponding ablation study, as shown in Figure~\ref{fig:abla_module}.
``w/o Reasoning Graph'' refers to a scenario where the multi-agent system generates outputs in natural language rather than using a structured reasoning graph. In this case, the F1 score decreased by 5.1\%, indicating that the construction of a reasoning graph enhances the analytical capability of the multi-agent framework in complex medical scenarios, leading to more accurate clinical reasoning. 
``w/o Multi-Agent'' refers to a situation where each agent constructs its reasoning graph independently and directly merges the results, without any inter-agent collaboration or update of the reasoning graph. In this case, the F1 score decreased by 6.1\%, suggesting that the reasoning graphs built by individual agents have limited coverage. Multi-agent collaboration allows for a more comprehensive analysis of key clinical evidence across different data sources, thereby improving the overall accuracy.
``w/o Referral Knowledge'' refers to a scenario where the expert agent does not refer to the clinical guideline when making decisions. In this case, the F1 score dropped by 2.3\%. The introduction of referral knowledge provides the expert agent with department transfer criteria and urgency guidance, enabling the agent to recommend the most appropriate department for a given medical scenario and thus improving the accuracy of medical referral decisions.

To investigate the impact of the four proposed relationship types (Cascade, Syndrome, Exclusion, and Context) on graph construction, we conduct ablation experiments as shown in Figure~\ref{fig:four_type_ablation}. Within the multi-agent framework, we explicitly remove selected relationship types to observe their influence on the final performance. Results demonstrate that the overall F1 score consistently increases as more relationship types are incorporated, indicating that the four defined relationship types align well with the characteristics of clinical reasoning and effectively support structured inference.
Furthermore, among all relationship types, Cascade exhibits the most substantial contribution to performance improvement. For example, in the “remove 1 relationship” setting, excluding Cascade leads to a drop in F1 from 95.3\% to 90.6\%. Similar significant declines are observed in both the ``remove 2'' and ``remove 3'' settings when Cascade is omitted. This finding suggests that, in complex medical referral scenarios, long-chain dependencies frequently exist across clinical evidence. By explicitly modeling these continuous inferential links through the proposed Cascade relation, our framework enhances the ability of multi-agent framework to perform coherent and clinically reliable reasoning.

\subsection{Impact of Multi-Agent Discussion Rounds}

We analyze the impact of the number of discussion rounds on the reasoning performance. In Figure~\ref{fig:multi_round}, the horizontal axis represents the number of discussion rounds (ranging from 1 to 5), and the vertical axis denotes the F1 score. We evaluate four representative lightweight foundation models as backbones within our framework: DeepSeek-V3, Qwen-Turbo, GPT-4o-mini, and Gemini-2.0-Flash. All models exhibit a performance increase as the discussion deepens from Round 1 to Round 3, demonstrating the effectiveness of the multi-agent collaboration mechanism. Among the evaluated models, DeepSeek-V3 achieves superior performance. It reaches a peak F1 score of 95.3\% at Round 3. Extending the discussion beyond this point (to Round 4 and 5) yields negligible gains or slight declines. This suggests that the reasoning graph converges effectively at Round 3, making it the optimal hyperparameter for efficiency and accuracy.

\subsection{Case Study}

\begin{figure*}[]
  \centering
  \includegraphics[width=0.95\linewidth]{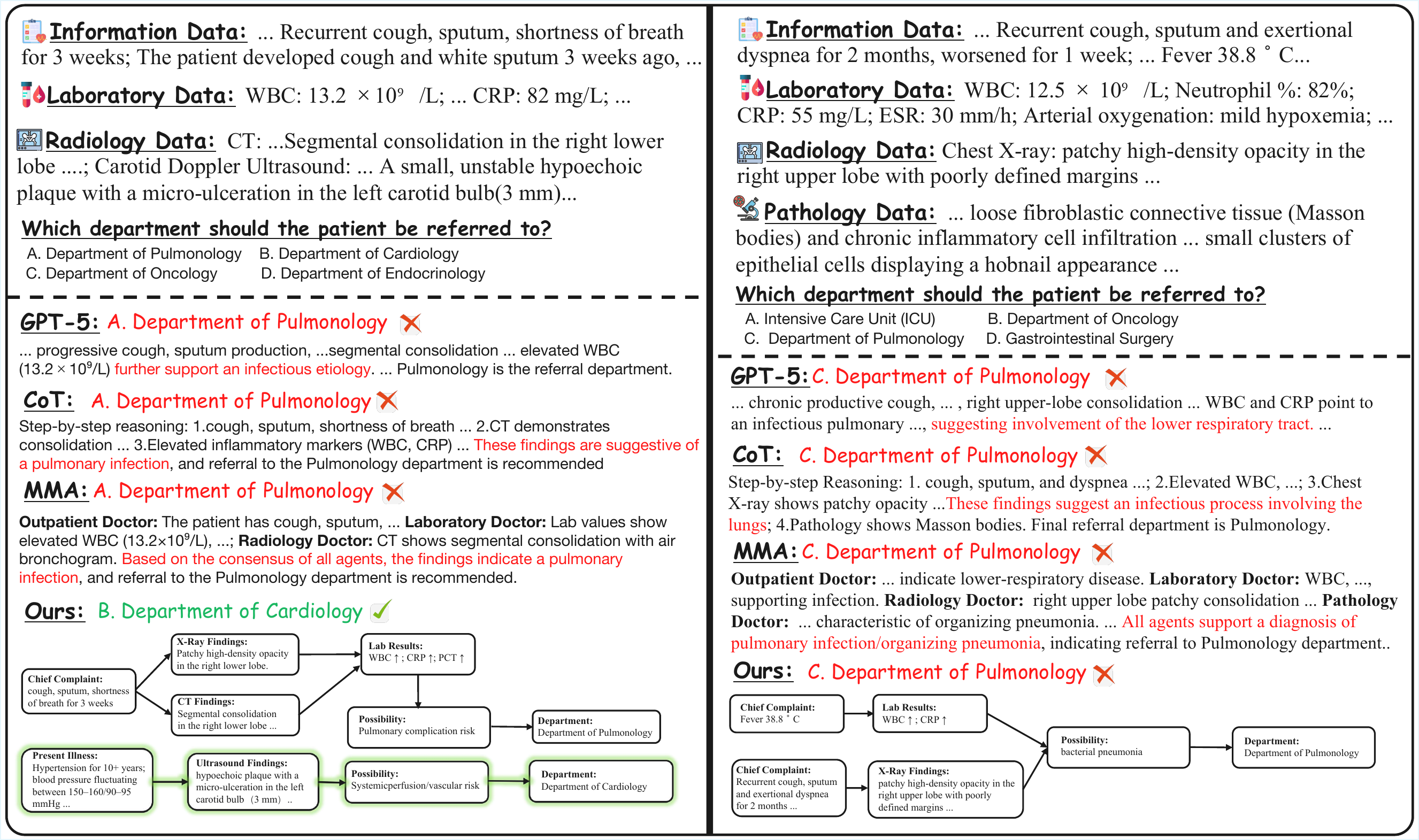}
  \caption{Qualitative comparison of case studies. The figure is divided into two parts: Case 1 (Left) and Case 2 (Right). In each part, the upper section presents the multi-source medical data, including patient information, laboratory results, radiology findings, and pathology data. The lower section displays the referral decisions and reasoning processes of GPT-5, CoT, MMA, and our method. Specifically, the bottom-most part illustrates the clinical reasoning graph constructed by our method to support its decision.}
  \label{fig:case_all}
\end{figure*}

To provide a concrete understanding of our method, we visualize two representative cases in Figure~\ref{fig:case_all}. The left part displays the first case, a patient presenting with recurrent cough, sputum, and shortness of breath. Methods like GPT-5, CoT, and MMA are misled by the prominent respiratory symptoms. They focus heavily on the ``cough'' and ``lung consolidation'' shown in the radiology data, leading them to incorrectly refer the patient to the Department of Pulmonology. In contrast, our method successfully identifies the underlying cause. By constructing a reasoning graph, our model captures critical non-respiratory evidence, specifically the ``10-year history of hypertension'' and the ``unstable plaque in the carotid bulb''. The graph connects these nodes to identify a high risk of systemic vascular failure. Consequently, our method correctly recommends the Department of Cardiology. The right part shows Case 2, which highlights a limitation of current agents. The patient presents with cough and fever, and pathology data shows ``Masson bodies'' (often a sign of inflammation) alongside epithelial cells with a ``hobnail appearance''. The correct referral is the Department of Oncology. However, GPT-5, CoT, MMA, and our method all incorrectly select the Department of Pulmonology. The models focus on the ``Masson bodies'' and inflammatory markers, interpreting the condition as simple pneumonia. They fail to recognize that the ``hobnail appearance'' is a high-risk feature for malignancy. In this case, the inflammation is likely a secondary change caused by tumor obstruction. Our agent lacks the specific expert knowledge to distinguish this subtle pathological feature. This suggests a need for future improvements, such as training more specialized sub-agents to detect fine-grained malignancy cues.

\section{Conclusion}
In this paper, we propose Multi-Agent Structured Graph Reasoning (MASGR), a multi-agent framework for complex medical referral task. In this framework, we first utilize four doctor agents to construct a reasoning graph which links the clinical evidences across different types of medical. Then, we set an expert agent combined with a pre-defined referral guideline to select the most appropriate department for the patient. Experiments show that MASGR outperforms other baseline methods, demonstrating stronger capability in cross-modal evidence interpretation and prioritizing referral decisions.

\section*{Acknowledgments}
This research is supported by the National Natural Science Foundation of China (62476097), the Fundamental Research Funds for the Central Universities, South China University of Technology (x2rjD2250190),  Guangdong Provincial Fund for Basic and Applied Basic Research—Regional Joint Fund Project (Key Project) (2023B1515120078), Guangdong Provincial Natural Science Foundation for Outstanding Youth Team Project (2024B1515040010), the Hong Kong Polytechnic University under the Postdoc Matching Fund Scheme (Project No. P0049003), the Hong Kong Research Grants Council under the Theme-based Research Scheme (project no. T22-501/23-R).

\printcredits

\bibliographystyle{cas-model2-names}

\bibliography{cas-refs}



\end{document}